\documentclass[11pt]{article}
\usepackage{amsmath}
\usepackage{cite}
\usepackage{epsfig}
\usepackage{fancyhdr}
\usepackage{float}
\usepackage[T1]{fontenc}
\usepackage{multirow}
\usepackage{shadow}
\usepackage{subfigure}
\usepackage{supertabular}
\usepackage{tabularx}

\begin{document}

\title{\textbf{Thermodynamics for single-molecule stretching
experiments}}
\author{J.M. Rubi,$^a$ D. Bedeaux$^b$ and S. Kjelstrup$^{b}$ \\
%EndAName
$^a$ Departament de Fisica Fonamental, Universitat de Barcelona,\\
Diagonal 647, 08028 Barcelona, Spain\\
$^b$Department of Chemistry, Faculty of Natural Science and Technology,\\
Norwegian University of Science and Technology, Trondheim, 7491-Norway}
\maketitle
\date{}

\begin{abstract}
\noindent We show how to construct non-equilibrium thermodynamics for
systems too small to be considered thermodynamically in a traditional sense.
Through the use of a
non-equilibrium ensemble of many replicas of the system which can be viewed
as a large thermodynamic system, we discuss the validity of non-equilibrium
thermodynamics relations and analyze the nature of dissipation in small
systems through the entropy production rate. We show in particular that the
Gibbs equation, when formulated in terms of average values of the extensive
quantities, is still valid whereas the Gibbs-Duhem equation differs from the
equation obtained for large systems due to the lack of the thermodynamic
limit. Single-molecule stretching experiments are interpreted under the
prism of this theory. The potentials of mean force and mean position, now
introduced in these experiments in substitution of the thermodynamic
potentials, correspond respectively to our Helmholtz and Gibbs energies.
These results show that a thermodynamic formalism can indeed be applied at
the single-molecule level. 


\end{abstract}

\section{Introduction}

Thermodynamics \cite{Callen} can only be applied to systems with an infinite
number of particles distributed in an infinite volume with a constant
density, a situation referred to as the thermodynamic limit. Under these
circumstances one may adopt a coarse description of the system ensuring the
existence of extensive quantities and justifying the absence of
fluctuations. The thermodynamic limit \cite{Reiss} can be asymptotically
approximated through statistical mechanics and the results from different
statistical ensembles converge. It is then said that thermodynamics deals
with large-scale systems.

The lack of a thermodynamic limit has important consequences for the
behavior of the system. Scaling down the size leads to a very different
scenario, in which contributions to the energy of the system not present in
the thermodynamic limit, such as surface energy, may show up, thereby
breaking extensivity and making a normal thermodynamic treatment impossible.
A typical example is a small cluster of $N$ particles whose Gibbs energy is
given by \cite{Hill}
\begin{equation}
G=\mu N+aN^{\beta }  \label{1}
\end{equation}%
with $\mu $ the chemical potential, $a$ an arbitrary function of the
intensive parameters and $\beta <1$ an exponent. When the cluster contains a
large number of particles, the second contribution becomes much smaller than
the first and can be neglected. At this limit one obtains the well-known
thermodynamic relation $G\cong \mu N$ for which the chemical potential is
the Gibbs energy per particle. On the other hand, for clusters containing a
smaller number of particles both contributions must be considered, which
makes the Gibbs energy non-extensive. As a consequence thermodynamics does
no longer apply. Without a coarse description, fluctuations become important
and play a role in the characterization of the system.

Despite this relevant feature, a thermodynamic treatment of small system is
still possible. The way to proceed was proposed by Hill \cite{Hill} for
systems at equilibrium and was subsequently applied to different situations such as  
biochemical cycle kinetics \cite{Hill2}, \cite{Qian1}, open metastable systems \cite{Hill.Chamberlin},
entropy-enthalpy compensation effects \cite{Qian2}, \cite{Qian3} and critical behavior 
of ferromagents \cite{Chamberlin}. It
was later referred to as nano-thermodynamics \cite
{Hill.nano}. Given a small system, the ensemble of many of its
replicas becomes a large system having the usual thermodynamic behavior. To
restore the form of the Euler equation, the expression of the Gibbs energy in terms of the chemical potential, one can express the Gibbs energy as 
\begin{equation}
G=\widehat{\mu }N  \label{2}
\end{equation}%
where the chemical potential $\widehat{\mu }$, according to Eq. (\ref{1}),
is given by \cite{Hill}
\begin{equation}
\widehat{\mu }=\mu +aN^{\beta -1}  \label{3}
\end{equation}%
Both chemical potentials thus coincide at the thermodynamic limit. It can also be
shown that thermodynamic quantities defined for different ensembles, may not
be equivalent \cite{Hill}.

Thermodynamics of small systems has been formulated at an equilibrium state 
\cite{Hill}. In many experimental situations, however, instead of being in a
quiescent state, the system evolves in time adopting different
non-equilibrium configurations. This situation is commonly found in kinetic
processes such as nucleation \cite{Reguera} and growth \cite{Gadomski} of
small clusters, in noncovalent association between proteins \cite{Qian4} and 
in active transport through biological membranes \cite{JTB}, \cite{PCCP}.
It is our purpose in this paper to develop the non-equilibrium
thermodynamics of small systems offering an interpretation of the
non-equilibrium thermodynamics concepts as well as identifying the role
played by dissipation.

The paper is organized as follows. Section 2 presents the main ideas of
Hill's thermodynamics of small systems. In Section 3 we proceed with the
study of small systems outside equilibrium by introducing the
non-equilibrium ensemble of replicas. The non-equilibrium thermodynamics
relations and the entropy production of small systems are obtained in
Section 4. In Section 5, we apply the non-equilibrium thermodynamics of
small systems developed to give a thermodynamic interpretation of
single-molecule experiments. Finally, in the conclusions section we
summarize our main results and discuss about perspectives of the theory
presented.

\section{The thermodynamic system of replicas}

A small system can be made large by constructing an ensemble of identical, independent
replicas. This procedure ensures the validity of thermodynamic relations in
the ensemble \cite{Hill}. Consider a group with K number of these replicas
with total energy $U_{tot}=K\overline{U}$, entropy $S_{tot}=KS$ and volume $%
V_{tot}=K\overline{V}$. Here $\overline{U}$ is the average internal energy, $%
S$ is\ the entropy and $\overline{V}$ the average volume per replica. The
total quantities are extensive by definition. According to the first law, we
can change the energy by adding heat or work to the ensemble, 
\begin{equation}
dU_{tot}=dQ+dW  \label{4}
\end{equation}%
At reversible conditions, the added heat $Q_{rev}$ follows from the
definition of the entropy 
\begin{equation}
dS_{tot}=\frac{dQ_{rev}}{T}  \label{5}
\end{equation}%
whereas the irreversible heat enters into the entropy production.

Let us assume that the system has a constant number of particles $N$ and is
in contact with pressure and heat reservoirs keeping its pressure $p$ and
temperature $T$ constant. Changes in the energy of the ensemble are given by 
\begin{equation}
dU_{tot}=TdS_{tot}-pdV_{tot}+\mu KdN+XdK  \label{6}
\end{equation}%
Here $\mu $ is the chemical potential of a large system of $N$ particles,
with $K=1$. The last term accounts for the energy raise by adding more
replicas to the ensemble and also represents an increase of entropy since
the total number of particles of the ensemble can spread over a larger
number of replicas. This term is not present in a large system for which $%
K=1 $.

The energy of the ensemble follows by integrating this expression, keeping $%
N $, $p$ and $T$ constant: 
\begin{equation}
U_{tot}=TS_{tot}-pV_{tot}+XK  \label{7}
\end{equation}%
and the average energy is in light of the relation between total and average
values given by 
\begin{equation}
\overline{U}=TS-p\overline{V}+G  \label{8}
\end{equation}%
where we have identified $X$ with the Gibbs energy $G$. Extensivity can be
restored by expressing this quantity as 
\begin{equation}
G=\widehat{\mu }N  \label{9}
\end{equation}%
with $\widehat{\mu }$ the corresponding chemical potential. Expression (\ref%
{8}) can also be written as 
\begin{equation}
\overline{U}=TS-p\overline{V}+\mu N+(\widehat{\mu }-\mu )N  \label{10}
\end{equation}

This equation does not correspond to the one derived for a thermodynamic
system. Deviations from the large-system behavior are due to the presence of
the term $(\widehat{\mu}-\mu)$ vanishing at the thermodynamic limit in which 
$G=\mu N$. Variations of the mean energy value are then given by 
\begin{equation}
d\overline{U}=TdS-pd\overline{V}+\mu dN  \label{11}
\end{equation}
showing that mean value variations obey the same thermodynamic relation as
for a large system.

A similar analysis can be performed for constant $N$, $V$ and $T$. One
obtains in this case 
\begin{equation}
dU_{tot}=TdS_{tot}-pKdV+\mu KdN+AdK  \label{12}
\end{equation}%
where $A=\overline{U}-TS$ is the Helmholtz energy which differs from $%
-pV+\mu N$, the value encountered in thermodynamics. The Gibbs equation
becomes 
\begin{equation}
d\overline{U}=TdS-pdV+\mu dN  \label{13}
\end{equation}%
The fact that the spreading of particles of the ensemble over the different
replicas depends on the conditions imposed on the system makes the
thermodynamic quantities for small systems different when these conditions
change. Thermodynamic relations thus depend on the ensemble used and consequently 
on the measurement techniques for the experiments.

\section{The statistical ensemble of non-equilibrium replicas}

Application of external forces or gradients move the system from the
reference state driving it through different non-equilibrium configurations
described by the set of coordinates $x$ and characterized by the probability
distribution function $P(x,t)$. The coordinates may for example represent
the number of particles of a cluster or the configurations of a polymer or
may in general be a reaction coordinate or an order parameter characterizing
the state of the system. The ensemble of K replicas becomes under the
presence of the external driving force a non-equilibrium ensemble whose
evolution in time is described by the probability distribution function $%
P(\Gamma ,t)$, with $\Gamma =(x_{1},..,x_{K})$ a point in the phase space of
the ensemble. Since the elements of the ensemble are by definition
independent systems, we have $P(\Gamma ,t)=P(x_{1})...P(x_{K})$. The
time-dependent entropy of the ensemble is, according to the Gibbs entropy
postulate, given by%
\begin{equation}
S_{tot}(t)=S_{tot,0}-k_{B}\int P(\Gamma ,t)\ln \frac{P(\Gamma ,t)}{%
P_{0}(\Gamma )}d\Gamma  \label{14}
\end{equation}%
where $S_{tot,0}$ is the entropy of the ensemble in the reference state, $%
P_{0}(\Gamma )$ the probability distribution of the ensemble in that state
and $k_{B}$ the Boltzmann constant. Using the fact that the replicas are
independent we can rewrite this equation as 
\begin{equation}
S_{tot}(t)=S_{tot,0}-K\left( k_{B}\int P(x,t)\ln \frac{P(x,t)}{P_{0}(x)}%
dx\right)  \label{15}
\end{equation}%
where $P_{0}(x)$ denotes the probability distribution of the system at the
reference state. The term within brakets of this expression represents the
change in the entropy of one replica system with respect to its value at the
reference state. Our construction then preserves the additive nature of the
non-equilibrium entropy.

\section{Non-equilibrium thermodynamics of small systems}

Based on the statistical expression for the entropy of the ensemble found in
the previous section, we will proceed to formulate the non-equilibrium
thermodynamics of small systems by deriving the relations common in
non-equilibrium thermodynamics and by giving the expression for the entropy
production.

\subsection{Thermodynamic relations}

We will assume that in the reference state, the system is at local
equilibrium. Changes in the non-equilibrium entropy during the evolution in
time of the ensemble, follows from the expression obtained for the
non-equilibrium entropy. We have 
\begin{equation}
dS_{tot}(t)=dS_{tot,0}-K\left( k_{B}\delta \int P(x,t)\ln \frac{P(x,t)}{%
P_{0}(x)}dx\right)  \label{16}
\end{equation}%
Entropy variations in the local equilibrium state have the same form as
those at equilibrium given through Eq.(1). Using this result in Eq.(6) one
obtains 
\begin{equation}
TdS_{tot}(t)=dU_{tot}(t)+pdV_{tot}(t)-N\widehat{\mu }dK-\mu _{0}KdN-K\left(
k_{B}T\delta \int P(x,t)\ln \frac{P(x,t)}{P_{0}(x)}dx\right)  \label{17}
\end{equation}%
where $\mu _{0}$ is the chemical potential of the system at the local
equilibrium state. The variations of the last term can be expressed as 
\begin{equation}
k_{B}T\delta \int P(x,t)\ln \frac{P(x,t)}{P_{0}(x)}dx=\int \mu
_{x}(x,t)\delta P(x,t)dx-\int \mu _{0}\delta P(x,t)dx  \label{18}
\end{equation}%
where we have defined the chemical potential along the $x$-coordinate as 
\begin{equation}
\mu _{x}(x,t)=k_{B}T\ln \frac{P(x,t)}{P_{0}(x)}+\mu _{0}  \label{19}
\end{equation}%
Using this last expression in (\ref{17}) together with the normalization of
the probability distribution 
\begin{equation}
N=\int P(x,t)dx\text{ \ \ }\rightarrow \text{ \ \ }dN=\int \delta P(x,t)dx
\label{20}
\end{equation}%
one arrives at the Gibbs equation of the ensemble 
\begin{equation}
TdS_{tot}=dU_{tot}+pdV_{tot}-N\widehat{\mu }dK-K\int \mu _{x}\delta Pdx
\label{21}
\end{equation}%
giving entropy changes at the local equilibrium state of the ensemble. We
suppressed the time and $x$ dependence of the variables to simplify the
notation. The last term integrates all entropy changes due to variations of
the configurations in $x$-space of the system. Using now the relations $%
S_{tot}=KS$, $U_{tot}=K\overline{U}$ and $V_{tot}=K\overline{V}$ and
equation (\ref{8}), we obtain from the Gibbs equation 
\begin{equation}
TdS=d\overline{U}+pd\overline{V}-\int \mu _{x}\delta Pdx  \label{22}
\end{equation}%
which shows that a Gibbs equation for small systems is valid when formulated
in terms of average values. The probability distribution function is the
average of the microscopic density distribution and can be expressed as $%
P(x,t)=<\delta (x-x(t))>$, where the average is taken over the possible
realizations of the Langevin force, for a given initial condition. This
probability is governed by a Fokker-Planck equation which for a
thermodynamic system can be derived within the framework of mesoscopic
non-equilibrium thermodynamics \cite{mnet}, \cite{Qian5}.

The Gibbs-Duhem equation relating variations of the intensive parameters can
be obtained from Eqs. (\ref{8}), (\ref{20}) and (\ref{22}) 
\begin{equation}
-SdT+\overline{V}dp-Nd\widehat{\mu}=\int(\widehat{\mu}-\mu_{x})\delta Pdx
\label{23}
\end{equation}
This equation differs from the one derived for large systems and tends to it
at the thermodynamic limit where $\widehat{\mu}$ and $\mu_{x}$ coincide. The sign of the integral depends on the type of process the system undergoes.

\subsection{Entropy production}

The Gibbs equation (\ref{22}) and the first law of thermodynamics,
formulated in terms of average values 
\begin{equation}
d\overline{U}=dQ-p_{ext}d\overline{V}  \label{24}
\end{equation}%
where $Q$ is the heat supplied and $p_{ext}$ the external pressure imposed
to the system, can be combined to obtain the rate of entropy change 
\begin{equation}
\frac{dS}{dt}=\frac{1}{T}\frac{d\overline{U}}{dt}+\frac{p}{T}\frac{d%
\overline{V}}{dt}-\frac{1}{T}\int \mu _{x}\frac{\partial P}{\partial t}dx
\label{25}
\end{equation}%
This equation can be rewritten in the well-known general form 
\begin{equation}
\frac{dS}{dt}=J_{s}+\sigma  \label{26}
\end{equation}%
where 
\begin{equation}
J_{s}=\frac{1}{T_{B}}\frac{dQ}{dt}  \label{27}
\end{equation}%
represents the entropy exchange with the environment, and $T_{B}$ is the
temperature of the thermal bath. The second contribution is the entropy
production rate 
\begin{equation}
\sigma =\left( \frac{1}{T}-\frac{1}{T_{B}}\right) \frac{dQ}{dt}+\frac{%
p-p_{ext}}{T}\frac{d\overline{V}}{dt}-\frac{1}{T}\int \ J\frac{\partial \mu 
}{\partial x}dx  \label{28}
\end{equation}%
which has to be non-negative due to the second law of thermodynamics.
To obtain this equation, in Eq. (\ref{25}) we have used the continuity
equation for the probability distribution function 
\begin{equation}
\frac{\partial P}{\partial t}=-\frac{\partial J}{\partial x},  \label{29}
\end{equation}%
with $J$ the probability current along the $x$-coordinate, and then
performed a partial integration neglecting boundary terms. As the
probability distribution, this current also represents an average value.

The first two contributions of the entropy production originate from the
interaction with the thermal bath due to the fact that the temperature and
the pressure of the system may not coincide with the corresponding
quantities of the bath. The third contribution results from the underlying
diffusion process along the $x$-coordinate. When this process is very fast,
the system reaches a quasi-equilibrium state almost immediately. In this
state the current does not depend on the coordinate and the last
contribution becomes proportional to $J\Delta \mu $. From the entropy
production one can now infer expressions for the rates, as is usually done
in non-equilibrium thermodynamics \cite{dGM}.

\section{Thermodynamics of single-molecule measurements}

Polymer stretching experiments have recently been performed to analyze the
validity of thermodynamics for measurements made on single molecules
arriving at the conclusion that when the system is too small the usual
thermodynamic relations are no longer valid and a thermodynamic treatment is
less useful \cite{Bustamante}. These experiments give different results when
the end-to-end distance of a single-molecule is held fixed and the force
fluctuates, compared to when the force is fixed and the end-to-end distance
fluctuates. These differences should vanish in the thermodynamic limit of an
infinite length. A single molecule is a particular case of a small system.
We shall show in this section that the theory we propose in this paper
provides a thermodynamic interpretation of the experimental measurements.
For this purpose, we will assume that the set of
measurements performed as a function of time 
defines the ensemble of replicas. As indicated by the experiments, the two
types of experiments must be analyzed using different environmental
variables. In both cases $N$ and $T$ are constant for the single molecule.
There are two ensembles: one in which the applied force is constant and the
other for which the end-to-end distance is constant.

\subsection*{Isometric experiments}

In the case in which the end-to-end distance $l$ and the temperature $T$ are
externaly controlled, energy variations in the ensemble of K replicas are,
in accordance with (\ref{12}), given by 
\begin{equation}
dU_{tot}=TdS_{tot}+fKdl+AdK  \label{30}
\end{equation}%
where the tensile applied force $f$ is interpreted as the change in the
internal energy of the ensemble with respect to the elongation of the
polymer per replica, keeping the entropy and the number of replicas
constant. In these experiments one employs a single replica and measures a
force which fluctuates as a function of the time. Averaging this quantity
over the time is equivalent to ensemble averaging and therefore gives the $f$
we need. Integration of this expression for a constant length leads to 
\begin{equation}
U_{tot}=TS_{tot}+AK  \label{31}
\end{equation}%
Using the definitions $U_{tot}=K\overline{U}$ and $S_{tot}=KS$ in this
relation we can identify $A=\overline{U}-TS$ with the Helmholtz energy
previously defined. $A$\ is the change of the internal energy due to
increasing the number of replicas by one, keeping $S_{tot}$\ and $l$\
constant. Variations of the average energy then follows from our previous
expressions 
\begin{equation}
d\overline{U}=TdS+fdl  \label{32}
\end{equation}

The Gibbs-Duhem equation can be derived from Eq. (\ref{32}) and the
definition of the Helmholtz energy, thus obtaining 
\begin{equation}
SdT+dA=fdl  \label{33}
\end{equation}%
If we now define the Helmholtz energy per unit of length 
\begin{equation}
\widehat{f}\equiv \frac{A}{l}  \label{34}
\end{equation}%
we obtain the Gibbs-Duhem equation 
\begin{equation}
-SdT+ld\widehat{f}=(f-\widehat{f})dl  \label{35}
\end{equation}%
From Eq. (\ref{33}) it follows that 
\begin{equation}
f=\left( \frac{\partial A}{\partial l}\right) _{T}  \label{36}
\end{equation}%
from which one finds the expression for the Helmholtz energy 
\begin{equation}
A(T,l)=\int_{0}^{l}f(T,l^{\prime })dl^{\prime }  \label{37}
\end{equation}%
where the integration constant is equal to zero. The Helmholtz energy is
therefore equal to the potential of mean force introduced by Keller et al. 
\cite{Bustamante}. It follows from this expression, that the quantity $%
\widehat{f}(l)$ is the average of the force $f(l)$ over the lengths between
zero and $l$ and is therefore generally different from $f$. This difference
is in relationship with its dependence on the polymer elongation. At the
thermodynamic limit of a very large $l$, $f\rightarrow \widehat{f}$. At this
limit, Eq. (\ref{35}) becomes the usual Gibbs-Duhem equation of
thermodynamics. 

In order to obtain the entropy production we will use the first law 
\begin{equation}
d\overline{U}=dQ+f_{ext}dl  \label{38}
\end{equation}%
In this equation the first term on the right hand side gives the heat which
enters the average replica from the heat bath. The second term is the work
done by an external force when one changes the length of the polymer, which
is the same for all replicas in the isometric experiments. Combined with Eq.
(\ref{32}) the first law yields the entropy rate of change 
\begin{equation}
\frac{dS}{dt}=\frac{1}{T}\frac{dQ}{dt}+\frac{1}{T}(f_{ext}-f)\frac{dl}{dt}
\label{39}
\end{equation}%
Considering that the temperature of the bath is equal to that of the system,
we can identify $\frac{1}{T}\frac{dQ}{dt}$ with the entropy flux into the
system and find for the average entropy production per replica 
\begin{equation}
\sigma =\frac{1}{T}(f_{ext}-f)\frac{dl}{dt}  \label{40}
\end{equation}%
In the isometric equilibrium experiments by Keller et al. \cite{Bustamante}
the length of the polymer is kept constant and $f_{ext}=f$. In that case the
entropy production vanishes. If one proceeds to change $f_{ext}$\ one can
describe the system using the linear force flux relation%
\begin{equation}
f_{ext}-f=\xi \frac{dl}{dt}  \label{41}
\end{equation}%
where $\xi $\ is the friction coefficient.

\subsection*{Isotensional experiment}

In the case in which the applied force $f$ and the temperature $T$ are
externaly controlled, energy variations in the ensemble of K replicas are
given by 
\begin{equation}
dU_{tot}=TdS_{tot}+fdl_{tot}+GdK  \label{42}
\end{equation}%
We define the average length of a replic by $l_{tot}=K\overline{l}$. The
Gibbs energy is 
\begin{equation}
G=\overline{U}-TS-f\overline{l}=A-f\overline{l}  \label{43}
\end{equation}%
In the thermodynamic limit $G=0$. Following the steps indicated in the
previous case, we obtain the Gibbs equation 
\begin{equation}
d\overline{U}=TdS+fd\overline{l}  \label{44}
\end{equation}%
and the Gibbs-Duhem equation 
\begin{equation}
-SdT-\overline{l}df=dG  \label{45}
\end{equation}%
from which we obtain the average elongation length 
\begin{equation}
\overline{l}=-\left( \frac{\partial G}{\partial f}\right) _{T}  \label{46}
\end{equation}%
In the experiment one keeps $f$ constant and measures $l$ as a function of
time. Averaging over time gives the ensemble average of $l$, which is equal
to the $\overline{l}$ we use. We conclude that 
\begin{equation}
G(T,f)=-\int_{0}^{f}\overline{l}(T,f^{\prime })df^{\prime }\equiv
X(T,f)  \label{47}
\end{equation}%
where $X(T,f)$ is the Gibbs energy in the presence of a force which
coincides with the potential of mean position introduced by Keller et al. 
\cite{Bustamante}. This quantity gives information about the mean elongation of the 
macromolecule for different values of the force.

In order to obtain the entropy production, we use the first law in the
following way 
\begin{equation}
d\overline{U}=dQ+f_{ext}\text{ }d\overline{l}  \label{48}
\end{equation}%
Proceeding as in the previous case, we obtain the entropy production 
\begin{equation}
\sigma =\frac{1}{T}\left( f_{ext}-f\right) \frac{d\overline{l}}{dt}
\label{49}
\end{equation}%
In the isotentional equilibrium experiments by Keller et al. \cite%
{Bustamante} $f_{ext}$ and $f$ were kept equal and constant while the
average length of the polymer was also constant. In that case the average
entropy production vanishes. If one proceeds to change $f_{ext}$\ one can
describe the system using the linear force flux relation%
\begin{equation}
f_{ext}-f=\xi \frac{d\overline{l}}{dt}  \label{50}
\end{equation}%
Comparing this equation to Eq.(%
\ref{41}) we see that while Eq.(\ref{41}) contains the average force $f$ and
the controlled length $l$, that Eq.(\ref{50}) contains the controlled force $%
f$ and the average length $\overline{l}$. In the thermodynamic limit both
equations become the same.

Calculating the Gibbs energy gives generally different results depending on
whether we use isometric or isotensional ensembles. As previously seen, when
the force is constant, $G=A-f\overline{l}$ whereas for a constant
elongation, we can construct $G$ from $A$ giving $A-\overline{f}l$. Only
when the force is linear in the elongation, or equivalently when the
Helmholtz energy is quadratic, both expressions coincide. Results obtained
from both of these ensembles would then be equivalent only when the force is
harmonic \cite{Bustamante}.

\section{Conclusions}

In this paper we have shown how non-equilibrium thermodynamics can be
constructed for the irreversible evolution of small systems. By considering
a non-equilibrium ensemble of many replicas as a large, and therefore
thermodynamic, system we can define extensive variables and then to use
well-known thermodynamic concepts and relations to apply these to small
systems. For different types of environmental variables one can derive the
equivalent Euler, Gibbs and Gibbs-Duhem equations and the entropy production
from which dissipation in small systems can be analyzed. The thermodynamic
relations obtained agree at the thermodynamic limit with the ones of
classical thermodynamics.

We have applied the formalism developed to interpret stretching experiments
performed with a DNA molecule. It has been argued that the thermodynamic
concepts and relations are less useful in these experiments and that the
potentials of mean force and mean elongation should replace the Helmholtz
and Gibbs energies \cite{Bustamante}. When we proceed along the lines
indicated, we can define extensive quantities and show that thermodynamics
does apply on the single-molecule level. We have shown that these potentials
introduced in the experiments coincide with our expressions for the
Helmholtz and Gibbs energies and that the thermodynamic relations depend on
the type of ensemble used. Our formalism gives the expression for the
entropy production corresponding to the experimental situations discussed.

Our analysis may provide a thermodynamic basis for experiments performed on
small systems which operate under non-equilibrium conditions for which the
absence of the thermodynamic limit and the importance of the fluctuations
may constitute relevant factors in the characterization of their equilibrium
and dynamical properties. Clusters, single molecules, small pumps 
and motors are cases where our theory
could ystematicaly be applied.

\subsubsection*{Acknowledgment}

JMR wishes to thank NTNU for financial support.

\pagebreak


\begin{thebibliography}{99}
\bibitem{Callen} Callen, H.B. Thermodynamics and an Introduction to
Thermostatistics, John Wiley and Sons, Inc.: New York, 1985.

\bibitem{Reiss} Reiss, H. Methods of Thermodynamics, Dover: New York, 1996.

\bibitem{Hill} Hill, T.L. Thermodynamics of small systems, Dover: New York,
1994.

\bibitem{Hill2}Hill, T.L. Free energy transduction and biochemical cycle kinetics,
Springer Verlag: New York, 1989.

\bibitem{Qian1} Qian, H. J. Phys.: Condens. Matter. \textbf{2005}, 17, S3783.

\bibitem{Hill.Chamberlin} Hill, T.L.; Chamberlin, R.V. Proc. Natl. Acad. Sci. USA 
\textbf{1998}, 95, 12779.

\bibitem{Qian2}Qian, H.; Hopfield, J.J. J. Chem. Phys. \textbf{1996}, 105, 9292.

\bibitem{Qian3}Qian, H. J. Chem. Phys. \textbf{1998}, 109, 10015.

\bibitem{Chamberlin} Chamberlin, R.V. Nature \textbf{2000}, 408, 337.

\bibitem{Hill.nano} Hill, T.L. Nano Lett. \textbf{2001}, 1, 111.

\bibitem{Reguera} Reguera, D; Rubi, J.M. J. Chem. Phys. \textbf{2001}, 115, 7100.

\bibitem{Gadomski} Gadomski, A.; Rubi, J.M. Chem. Phys. \textbf{2003}, 293, 169.

\bibitem{Qian4} Qian, T.L. J. Math. Biol. \textbf{2006}, 52, 277. 

\bibitem{JTB} Kjelstrup, S.; Rubi, J.M.; Bedeaux, D. J. Theor. Biol. \textbf{2005}, 234, 7.

\bibitem{PCCP} Kjelstrup, S.; Rubi, J.M.; Bedeaux, Phys. Chem. Chem. Phys. \textbf{2005}, 7 4009.

\bibitem{mnet} Reguera, D.; Rubi, J.M.; Vilar, J.M.G. J. Phys. Chem. B \textbf{2005},
109, 21502.

\bibitem{Qian5} Qian, H. Phys. Rev. E \textbf{2001}, 65, 016102.

\bibitem{dGM} de Groot, S.R.; Mazur, P. Non-Equilibrium thermodynamics,
Dover: New York, 1984.

\bibitem{Bustamante} Keller, D.; Swigon, D.; Bustamante, C. Biophys. J.
\textbf{2003}, 84, 733.

\end{thebibliography}
\end{document}